# Quantum Cascade Laser Based Hybrid Dual Comb Spectrometer

LuigiConsolino,[1,*] MalikNafa,[1] MicheleDeRegis,[1] FrancescoCappelli,[1] KatiaGarrasi,[2] FrancescoP.Mezzapesa,[2] LianheLi,[3] A.GilesDavies,[3] EdmundH.Linfield,[3] MiriamS.Vitiello,[2] SaverioBartalini[1,4] and PaoloDeNatale[1]

[1] CNR-Istituto Nazionale di Ottica and LENS, Via N.Carrara 1, 50019 Sesto Fiorentino (FI), Italy
[2] NEST, CNR - Istituto Nanoscienze and Scuola Normale Superiore, Piazza S. Silvestro 12, 56127, Pisa, Italy
[3] School of Electronic and Electrical Engineering, University of Leeds, Leeds LS2 9JT, UK
[4] ppqSense Srl, Via Gattinella 20, 50013 Campi Bisenzio FI, Italy

[*] luigi.consolino@ino.it

**Four-wave-mixing-based quantum cascade laser frequency combs (QCL-FC) are a powerful photonic tool, driving a recent revolution in major molecular fingerprint regions, i.e. mid- and far-infrared domains. Their compact and frequency-agile design, together with their high optical power and spectral purity, promise to deliver an all-in-one source for the most challenging spectroscopic applications. Here, we demonstrate a metrological-grade hybrid dual comb spectrometer, combining the advantages of a THz QCL-FC with the accuracy and absolute frequency referencing provided by a free-standing, optically-rectified THz frequency comb. A proof-of-principle application to methanol molecular transitions is presented. The multi-heterodyne molecular spectra retrieved provide state-of-the-art results in line-center determination, with the same accuracy levels of nowadays available molecular databases. The devised setup provides a solid platform for a new generation of THz spectrometers, paving the way to more refined and sophisticated systems exploiting full phase control of QCL-FCs, or Doppler-free spectroscopic schemes.**

## Introduction

High-precision molecular spectroscopy in the terahertz range of the electromagnetic spectrum (0.1-10 THz) has attracted, in the last decades, ubiquitous attention from the scientific community. Indeed, it promises to enable fundamental physics research (e.g. time variation of fundamental constants,[1,2] assessment of the electric dipole moment of the electron,[3] parity violation in molecules[4]) while simultaneously providing deep impact on real-life applications,[5] spanning from environmental monitoring[6,7] to health science,[8,9] from communication[10,11] to security.[12,13] Actually, many scientifically intriguing molecules present characteristic rotational and ro-vibrational absorptions in the THz region. Therefore, this domain is nowadays considered as a fingerprint region still under development.[14] As a consequence, one of the big photonic challenges of the last decades has been the realization of THz laser sources capable of addressing absorption linestrengths comparable or even stronger than fundamental, mid-infrared (mid-IR), vibration transitions, but over much narrower Doppler-limited linewidths, ruled by inverse linear relationship with the wavelength.[15] In order to tackle metrological-grade spectroscopic applications, these lasers sources should ideally be broadband or widely tunable, spectrally pure, high-power, and traceable against a primary frequency standard.[16] To this purpose, THz sources based on cascaded frequency multiplication[17,18] or difference frequency generation sources[19,20] can be used, although they usually emit low (µW) power levels. Single-frequency THz quantum cascade lasers (QCLs) have recently shown metrological-grade performances,[21] at the cost of limited mode-hop-free tunability range (few hundred MHz). Instead, spectroscopic setups based on frequency combs (FCs) offer the possibility to overtake much of these disadvantages.

In the frequency domain, a FC consists of a series of coherent and equally spaced lines, with a fixed phase relation, covering a broad spectrum. The coherence of the optical modes allows all the modes of a FC to be fully controlled in phase by only two parameters: repetition and carrier offset frequencies.[22] Initially conceived for frequency metrology,[23–25] FCs have been straightforwardly applied to high-precision spectroscopy of samples, with a variety of different techniques.[26,27] The importance of using frequency combs for broadband spectroscopy relies on the exploitation of the frequency accuracy and spectral resolution provided by stabilized FC modes. However, this happens only if each comb mode is spectrally resolved, otherwise the comb acts only as a bright, broadband light source. For this reason, a number of high-resolution dispersive as well as Fourier transform spectrometers have been developed, capable of resolving individual comb modes.[27] Among these techniques, one of the most successful is multi-heterodyne spectroscopy, commonly known as dual comb spectroscopy (DCS),[28,29] which allows retrieving Fourier spectra without the need of moving interferometric components. This translates in much faster and spectrally-resolved acquisitions. In the most general picture, a DCS setup consists of two FCs with slightly different repetition frequencies, mixed onto a fast detector. This generates a down-converted radio-frequency comb, whose modes carry all the relevant information of the mixed FCs. In order to acquire molecular spectra, one or both FCs interact with a gaseous sample before frequency mixing. In this way, the sample spectral information is encoded into the FCs radiation, and therefore translated to the radio frequency (RF) domain, where it can be easily acquired and analyzed.

The possibility of migrating FC sources, and therefore DCS setups, in the THz region has been initially allowed by down-converted frequency combs (DC-FCs), obtained by down-conversion of visible or near-IR pulses in photoconductive antennas[30] or non-linear crystals[31]. Since then, a number of different THz dual-comb spectrometers have been developed (see ref.[32] and references therein). The main drawback of these setups, however, is the µW level power provided by the small efficiency of the involved non-linear processes. Recently, a new approach to THz FCs has been proposed, which relies on QCLs,[33,34] representing one of the most significant developments for semiconductor physics of the last 25 years.[35]

QCLs are current-driven semiconductor lasers based on intersubband transitions, whose spectral emission can be engineered with a proper nanoscale tailoring of the quantum wells building up the laser heterostructure. With respect to non-linear down-converting setups, they emit much higher power coherent THz radiation, even though their widespread use is still hampered by the need of cryogenic cooling.[36,37] Moreover, the short upper lasing state lifetime, compared to the cavity round trip time, prevents classical passive mode-locking in QCLs.[38] Active pulsed mode-locking has been achieved in THz QCLs, with limitations on pulses duration that cannot match the inverse of the gain bandwidth, which, by itself, limits significantly pulse shortening at values as low as 4 ps.[39,40] Nevertheless, a different route to THz FCs based on QCL-FCs has been demonstrated, in the mid-IR[41] and THz[42] regions, based on multimode-emitting, broad-gain Fabry-Pérot devices, that can spontaneously achieve comb operation regime, over a specific range of biases, thanks to four-wave-mixing (FWM) non-linear effects[43]. Thorough characterization of these devices with shifted-wave interference Fourier-transform spectroscopy (SWIFTS)[44–46] and Fourier-transform analysis of comb emission (FACE)[47,48] techniques has proven that, thanks to FWM, a tight and non-trivial phase relation is established among the emitted modes, leading to an emission profile which is deeply frequency and amplitude modulated, rather than pulsed.

QCL-FCs can therefore be used in DCS setups, aiming to exploit their high (>10 mW) optical power. This has been demonstrated in the mid-infrared[29] but, in this regard, the THz region is still lagging behind. In fact, an etalon signal (i.e. simulating a real molecular absorption) has been recorded in 2016,[49] and, more recently, low resolution spectra of ammonia gas[50] and water vapour have been acquired.[51,52] With respect to DC-FCs-based setups, one of the limiting factors is the lack of mutual coherence between the two QCL-FCs, which leads to much shorter integration times, and, more importantly from a metrological point of view, to the lack of suitable referencing to the primary frequency standard.

Here, we demonstrate a hybrid approach to THz DCS, capable of merging the two main advantages of DC-FCs and QCL-FCs based dual-comb spectrometers, i.e. the accurate and absolute referencing of the retrieved molecular absorption frequency on one side, and the high sensitivity ensured by the high emitted power on the other. We believe that the proof-of-principle demonstration of this novel approach can finally unleash the full potential of THz DCS setups. Our spectrometer has been characterized and tested on molecular absorption profiles of methanol vapours, returning a state-of-the-art absolute accuracy of about $5 \cdot 10^{-8}$ on the measured transitions line center determination.

## Results

**Experimental Setup**

The THz hybrid dual-comb spectrometer setup is based on multi-heterodyne down-conversion of a THz QCL-FC and a fully-stabilized optically rectified THz frequency comb (OR-FC), generated by means of a femtosecond telecom laser (Menlo Systems, FC1500) and a non-linear crystal waveguide.[31] A schematic illustration of the setup is presented in fig. 1a. A full description of the employed high dynamic range QCL-FC can be found in ref.[53] It is mounted on the cold finger of a liquid helium flux cryostat, and is driven in continuous-wave (CW) operation by an ultra-low-noise current driver (ppqSense, QubeCL-P05). Its repetition frequency, corresponding to the inverse of the cavity round trip time, is about 17.45 GHz, and can be extracted as intermodal beatnote frequency ($f_{IBN}$), by means of a bias-tee (Marki Microwave, BT-0024SMG), mounted very close to the device. The same bias-tee can be also alternatively used for injection locking of the QCL-FC spacing, by means of a local oscillator ($f_{LO}$). The QCL-FC output beam is collimated by means of an off-axis parabolic mirror, and propagates through a spectroscopy cell, filled with methanol vapours at a selectable pressure. The OR-FC has a repetition rate tunable by ~2% around 250 MHz ($f_{rep}$), and covers a broad spectrum of about 7 THz. For the purposes of this setup, it is optically filtered close to the center frequency of the QCL-FC, at around 3 THz. The emitted OR-FC beam is then fully transmitted through an *ad-hoc* oriented wire-grid polarizer (WGP), that also acts as beam coupler. In fact, the QCL-FC beam is superimposed to the OR-FC beam using the WGP, and its transmittance/reflectance ratio is selected by means of a quarter waveplate (λ/4). The superimposed FC beams are then coupled to a fast mixer, i.e. an He-cooled hot electron bolometer (HEB, Scontel Technologies, RS0.3-3T1) that performs the down-conversion to radio frequencies. The radio frequency range, down-converted frequency comb (RF-FC) is acquired by a spectrum analyzer (Tektronix, RSA5106A) with a 40 MHz real-time bandwidth. The portion of the QCL-FC that is not used for the multi-heterodyne down-conversion, representing most of the QCL emission, is coupled to a pyroelectric detector for QCL power monitoring, and for direct absorption spectroscopy. A GPS-disciplined Rubidium-Quartz oscillator chain is used as common frequency standard for the OR-FC repetition rate stabilization, for the spectrum analyzer acquisitions, and for the local oscillator $f_{LO}$.

An example of the RF-FC spectrum acquired by the spectrum analyzer is presented in fig. 1b. In the down-conversion process the OR-FC repetition frequency is tuned to have a quasi-integer ratio between the two. This configuration makes the RF-FC heterodyne beatnotes (HBNs) distinguishable while sufficiently close, so that the spectroscopic information encoded in the QCL-FC light can be easily retrieved. Figure 1b shows 12 modes retrieved in the down conversion process, but while the ones close to the edge of the spectral coverage present a low signal to noise ratio, the most intense 9 modes can be used for molecular interrogation. It can also be noted that in the acquisition timescales their linewidth is about 200-500 kHz, depending on the mode order.

**Hybrid Dual Comb Spectroscopy Acquisitions**

The QCL-FC emitted frequencies can be simultaneously tuned by changing either the device driving current, its operational temperature, or, in case of active injection locking loop, by changing the local oscillator frequency $f_{LO}$. This latter possibility does not add, at the moment, any improvement in terms of spectroscopic accuracy, but will be an interesting feature in the next future, as discussed in the next section. While performing these scans, we acquire both the direct absorption and the DCS signals. The former is acquired by means of a pyroelectric detector and, to this purpose, the QCL beam is first modulated by a chopper wheel (see fig 1a), then demodulated using a lock-in amplifier. The DCS signal is obtained by acquiring a sequence of RF-FC spectra, and then extracting the amplitude of each HBN signal. A schematic representation of the acquisition procedure is reported in fig 2. In this way, we simultaneously scan and acquire portions of the molecular gas sample spectrum, that are separated by the QCL-FC intermodal frequency ($f_{IBN}$), along the whole QCL-FC spectral coverage. In the specific case of the device employed in this work we scan 9 portions of the THz spectrum, corresponding to the 9 most intense QCL-FC modes. Thanks to the frequency scale provided by the OR-FC, the RF frequencies of the HBN can be readily converted into absolute THz frequencies, as thoroughly explained in the Methods section. It is worth noticing that the proposed setup is also robust with respect to experimental instabilities (mechanical vibrations, current and temperature fluctuations). In fact, these are translated into frequency variations of the QCL-FC modes, that are taken into account thanks to the absolute frequency scale.

**Spectroscopic Measurements**

In this work, the hybrid THz dual comb spectrometer has been tested on methanol vapours at low pressure, and the QCL-FC frequency scans are performed at a fixed heat sink temperature of 29.5 K, while the device current is tuned in a 15 mA current range, with steps of about 0.05 mA. Figure 3a shows the direct absorption spectroscopy signal acquired with a methanol gas pressure of 170 Pa, while scanning the QCL driving current. Two absorption profiles are clearly visible but, as this pyroelectric signal is an integral of the intensities of all the QCL-FC modes, it is possible neither to distinguish which modes are involved in the absorption, nor to retrieve the absolute frequencies of the transitions. At the same time fig. 3b and 3c show portions of the DCS signal related to the mode labeled as 11 in fig 1b. In fact, the two absorption profiles are attributed to this mode only, while no other transitions are probed by the other modes during the scan. It is worth noticing that all the other modes' frequencies are also simultaneously scanned and, if there were any transition, it would have been acquired and measured by our spectrometer. The absolute frequency scale calibration procedure, described in the Methods section, is applied for QCL-comb mode 11 close to both transition linecenters, and the resulting absolute frequency is shown on the x-axis of fig. 3b and 3c. From Voigt fits of experimental data, single-shot acquisition accuracy of few hundreds kHz is obtained, i.e. a relative accuracy of $\sim 7 \cdot 10^{-8}$ on a single measurement.

Thanks to the absolute frequency scale, the two methanol transitions have been identified as the $(J, K) = (11, 11 \rightarrow 12, 12)$, and the $(J, K) = (12, 11 \rightarrow 13, 12)$, where J and K are the quantum numbers identifying the total angular momentum and its projection on the molecule axis. From here on these transitions are reported as transitions T1 and T2, respectively. In order to allow comparison of the absolute center frequency with molecular databases, the self pressure shift of each transition has to be measured. For this reason the two molecular profiles were probed and acquired at different methanol gas pressure. Figure 4 shows acquisitions of the T1 methanol transition around 3.105363 THz. The experimental data and the Voigt fits are shown at gas pressures of 100, 140 and 240 Pa (fig. 4a, 4b and 4c respectively), while the last panel shows the retrieved self pressure shift. The center frequency of the transition, and the uncertainty affecting the measurement can be calculated by the linear fit shown in fig. 4d. The frequency value extrapolated at zero gas pressure is 3105363072(141) kHz, corresponding to an accuracy of $4.5 \cdot 10^{-8}$. The less intense transition shown in fig. 3 corresponds to the T2 methanol transition, acquired at pressures of 100, 180 and 300 Pa, respectively. These acquisitions, together with the retrieved pressure shift, are shown in fig. 5. The center frequency extrapolated at zero pressure is 3105511610(196) kHz, corresponding to a relative uncertainty of $6.3 \cdot 10^{-8}$. The results of the two linear fits are summarized in Table 1.

Finally, fig. 6 shows the transitions center frequencies retrieved in this work, compared with previously available data for the same transitions. In particular, we report comparison with data from the Jet Propulsion Laboratory Molecular Spectroscopy (JPL) Catalog, by the California Institute of Technology, based on the simulations by Xu et al,[54] with other experimental measurements,[55] and with calculated values from the same authors. In this latter work the center frequency values were measured at 200 Pa gas pressure, while the self-pressure shift was not retrieved. For this reason we also report experimental data of ref.[55], frequency shifted according to our measured pressure shift coefficients of -2.8(0.8) kHz/Pa for T1, and -4.7(1.0) kHz/Pa for T2 (see Table 1). Remarkably, the comparison shows that our spectrometer allows performing measurements with the same level of accuracy of state-of-the-art molecular databases (JPL database), while its accuracy improves previous experimental measurements on the same transitions by one order of magnitude. Moreover the measurements on both transitions T1 and T2 are in full agreement with the molecular database, while we have to make a distinction when comparing to ref.[55]. In fact, regarding transition T1, fig. 6a shows that the agreement of our measurements with the values reported in the JPL catalog is very good, but these both disagree with calculated and experimental values reported in ref.[55]. The self-pressure shift coefficient alter these latter data in the right direction, but not enough to provide agreement. Regarding transition T2 (fig. 6b), while the agreement between our data and the JPL catalog is good, there is good agreement also with the other experimental value. Once again, the measured pressure shift coefficient improves the agreement on experimentally measured values.

## Discussion

We have devised a THz hybrid DCS setup capable of performing molecular multi-heterodyne spectroscopy with state-of-the-art accuracy, challenging existing molecular databases. The proof-of-principle demonstration of frequency scans over the wide spectral coverage of a a state of the art, high dynamic range QCL-FC, at this level of accuracy, can pave the way to a new generation of trace gas analyzers based on THz radiation. In fact, while the transitions probed in this work were probed by the same mode, it is easy to imagine an *ad-hoc* designed QCL-FC capable of simultaneously probing different transitions of a molecular gas, or even transitions of different molecular samples in a low-pressure gas mixture.

The spectrometer, however, presents some limitations that can be overcome in the next future. The limited tunability of the modes of the QCL device with temperature/current scans, that in our case is of few hundreds MHz, can be more than enough to perform high-accuracy acquisitions on selected molecular transitions, but might be a limitation in the analysis of complex gas mixtures. For these challenging applications, our spectrometer can in principle be equipped with an array of slightly frequency detuned THz QCLs, as well as broader dynamic range devices, such as dispersion compensated lasers,[56] or saturable-absorbers-equipped QCL-FCs. These devices can provide a broader spectral coverage and, at the same time, FC operation at higher driving currents, providing higher output power and broader frequency scans. Finally, the possibility of using room temperature difference frequency generation THz QCLs[57] can move our spectrometer from table-top laboratory size to *in-situ* operation size.

The accuracy on retrieval of linecenter frequencies is limited by the signal to noise ratio of the spectroscopic acquisitions. This limitation can be readily overcome by performing acquisitions with a fully phase-locked QCL-FC.[48] In fact, the possibility to perform frequency scans by tuning both the device current and the local oscillator frequency with active injection locking, confirms the feasibility of these acquisitions, and provides complete and independent control on the frequencies of the individual QCL-FC modes. This mode of operation was demonstrated for direct absorption spectroscopy,[21,58] and we therefore expect to improve the accuracy of our spectrometer by one order of magnitude, shifting the spectroscopic accuracy limit to the Doppler broadened transition profiles acquired. This further limitation can be overcome using sub-Doppler spectroscopic techniques, that are precluded to DC-FCs-based DCS, but can be implemented in our spectrometer, thanks to the significantly higher output power provided by a QCL-FC. Indeed, saturation effects and pump and probe setups have been implemented with QCL-based spectrometers,[59,60] although no metrological-grade measurements have been obtained yet. Finally, although having low finesse values, newly-developed THz resonant cavities[61,62] can also help to further enhance the spectrometer sensitivity, as routinely done in the visible or near infrared spectral domains.[63]

# Methods

**Absolute frequency scale calibration**

The absolute frequency of each and every QCL-FC emitted mode can be retrieved thanks to the absolute frequency scale provided by the phase stabilized OR-FC, as schematically depicted in fig. 7. The frequency of the $N^{th}$ QCL-FC mode can be written as:

$$f_N = M f_{rep} \pm f_{b,N} \qquad \text{eq. 1}$$

where $M$ is the order of the OR-FC mode beating with the $N^{th}$ QCL-FC mode at higher or lower frequency, corresponding to different signs of the HBN frequency $f_{b,N}$. As thoroughly explained in ref. [64], by keeping the QCL mode $N$ frequency constant (i.e. keeping constant both operational temperature and driving current), by tuning $f_{rep}$ and acquiring $f_{b,N}$, the order $M$ of the OR-FC mode can be unambiguously determined by a simple linear fit of eq. 1. Knowing the intermodal beatnote frequency $f_{IBN}$, this also provides an absolute frequency scale for all the emitted QCL-FC modes. Indeed, the absolute frequency of each mode can be reconstructed by simply adding to the frequency of the $N^{th}$ mode $f_{IBN}$ (or $f_{LO}$ in case of active injection locking), times the order difference.

This procedure has been carried out twice, while the QCL is close to resonance with transitions T1 and T2, respectively. The linear fits retrieving the order of the OR-FC mode beating with QCL-FC mode 11 are presented in fig. 8. The outcome of the fit regarding the slope are then rounded to the nearest integer, i.e. the order $M$ of the OR-FC. Once the absolute frequency scale of the QCL-FC modes is calibrated, any variations in their frequencies, due to current or temperature tuning, as well as to experimental instabilities, is translated into exactly the same frequency variations of the corresponding HBN of the RF-FC. In this way it is possible to constantly track and monitor all the emitted QCL-FC frequencies simultaneously.

**Data availability:** The data supporting the findings of this study are available from the corresponding author upon reasonable request.

**Acknowledgements**

The authors acknowledge financial support from the Qombs Project, EU-H2020-FET Flagship on Quantum Technologies grant no. 820419, the ERC Project 681379 (SPRINT), the Italian ESFRI Roadmap (Extreme Light Infrastructure-ELI), EC–H2020 Laserlab-Europe grant agreement 654148. We also acknowledge support from the EPSRC (HyperTerahertz programme, EP/P021859/1). EHL acknowledges support from the Royal Society and Wolfson Foundation.

**Authors contributions**

L.C. and S.B. conceived the experiment, L.C., M.N. and M.D.R. performed the measurements and analyzed the data, L.C. wrote the manuscript, M.N., M.D.R., F.C., M.S.V. and P.D.N. contributed to manuscript revision, K.G., F.P.M. and M.S.V. fabricated the THz QCL frequency comb, L.L., A.G.D. and E.H.L. grew the QCL heterostructure, L.C., M.N., M.D.R. and F.C. discussed the results, M.S.V. supervised the design and development of the THz QCL comb, the work was done under the joint supervision of P.D.N. and L.C..

**Competing interests:** The authors declare no competing interests.

# Figure Captions

Figure 1 : Hybrid Dual Comb Spectrometer. **a** Experimental setup scheme. OR-FC: Optically Rectified Frequency Comb; QCL-FC: Quantum Cascade Laser Frequency Comb; WGP: Wire Grid Polarizer; HEB: Hot-Electron Bolometer; SA: Spectrum Analyzer; BP filter: band-pass filter. **b** Acquisition of the heterodyne beatnotes signal (HBN) on the SA, resulting from the mixing of the OR- and QCL-FC on the HEB.

Figure 2: Schematic representation of the hybrid dual-comb spectroscopy signal acquisition. Two quantum cascade laser frequency comb (QCL-FC) modes N (red) and M (blue) spaced by $(N-M) \cdot f_{IBN}$ are frequency tuned by driving current or heat sink temperature variations. This tuning corresponds to variations in the down-converted heterodyne beatnotes (HBN) signals, having amplitudes that are continuously monitored. Retrieval of the variations in the HBN amplitudes enables simultaneous reconstruction of molecular transitions that are spaced by $(N-M) \cdot f_{IBN}$ in the THz frequency domain. RF: radio frequency; $I_{QCL}$: QCL driving current; $T_{QCL}$: QCL heat-sink temperature.

Figure 3: Direct absorption and multi-heterodyne dual-comb spectroscopy (DCS) signals. **a** Direct absorption spectroscopic signal acquired from the pyroelectric detector, as integral of all the simultaneously emitted quantum cascade laser frequency comb (QCL-FC) modes. Two absorption features are observed at different QCL driving current. **b** and **c** Multi-heterodyne DCS spectra corresponding to these absorption features. Each point corresponds to the amplitude of the heterodyne beatnote (HBN) involved in the absorption, acquired in the down-converted radio frequency frequency comb spectrum. In these spectra the HBNs present a flat top, due to frequency instabilities. As a consequence, the uncertainty bars have been calculated as standard deviation on all the points of the HBN flat top.

Figure 4: Absorption coefficient (α) acquired from the multi-heterodyne dual comb spectroscopy spectra of the transition (J, K) = (11, 11 → 12, 12). The transition has been acquired at methanol pressures of **a** 100, **b** 140 and **c** 240 Pa. Each point corresponds to the amplitude of the heterodyne beatnote (HBN) involved in the absorption, acquired in the down-converted radio frequency frequency comb spectrum. In these spectra the HBNs present a flat top, due to frequency instabilities. As a consequence, the uncertainty bars have been calculated as standard deviation on all the points of the HBN flat top. **d** The measured pressure shift is reported in the last panel, corresponding to a slope of -2.8(0.8) kHz/Pa. The retrieved center frequency for this transition is 3105363072(141) kHz. The uncertainty on the x-axis comes from the pressure gauge instrumental limit, while the uncertainty on the frequency center is due to the precision of the Voigt fits shown in the previous panels.

Figure 5: Absorption coefficient (α) acquired from the multi-heterodyne dual comb spectroscopy spectra of the transition (J, K) = (12, 11 → 13, 12). The transition has been acquired at methanol pressures of **a** 100, **b** 180 and **c** 300 Pa. Each point corresponds to the amplitude of the heterodyne beatnote (HBN) involved in the absorption, acquired in the down-converted radio frequency frequency comb spectrum. In these spectra the HBNs present a flat top, due to frequency instabilities. As a consequence, the uncertainty bars have been calculated as standard deviation on all the points of the HBN flat top. **d** The measured pressure shift is reported in the last panel, corresponding to a slope of -4.7(1.0) kHz/Pa. The retrieved center frequency for this transition is 3105511610(196) kHz. The uncertainty on the x-axis comes from the pressure gauge instrumental limit, while the uncertainty on the frequency center is due to the precision of the Voigt fits shown in the previous panels.

Figure 6: Comparison between our measured linecenters (orange) for the transitions **a** (J, K) = (11, 11 → 12, 12) and **b** (J, K) = (12, 11 → 13, 12), and literature values retrieved from: the Jet Propulsion Laboratory (JPL) catalog (blue), measurements at 200 Pa by Moruzzi et al.[55] (green), the same values frequency shifted thanks to the self-pressure shift coefficient measured in this work (red) and calculation performed by the same authors (violet). The error bars on our measurements come from the linear fit extrapolating the linecenter frequencies at zero pressure.

Figure 7: Illustration of the multi-heterodyne down-conversion process. Schematic representation of the quantum cascade laser (QCL-) (red) and optically rectified (OR-) frequency comb (FC) (blue), whose modes are respectively spaced by $f_{IBN}$ and $f_{rep}$. These two repetition frequencies are tuned close to an integer ratio, allowing an ordered and distinguishable down-conversion to radio frequencies (RF). In fact, the down-converted RF-FC modes (green) are equally spaced by D, and their easily measurable RF frequencies are used to calibrate the absolute frequency scale of the QCL-FC, as described in the main text.

Figure 8: Retrieval of the optically rectified frequency comb (OR-FC) modes order. The frequency of the beatnote $f_b$ is plotted as a function of the OR-FC repetition frequency, while keeping the quantum cascade laser (QCL) driving current and operational temperature fixed. Plots **a** and **b** are relative to QCL mode 11 frequency interacting with transitions T1 and T2 respectively. The order $M$ of the linear fit is extracted as linear slope, and then rounded to the nearest integer, i.e. $M$=12455 for panel **a** and $M$=12456 for panel **b**.

Table 1: Self-pressure shifts retrieval. The outcome of the linear fits reported in fig. 4d and fig. 5d are summarized in this table. The retrieved intercept value identify the zero-pressure transition frequency, while the slopes correspond to the self-pressure shift parameters.

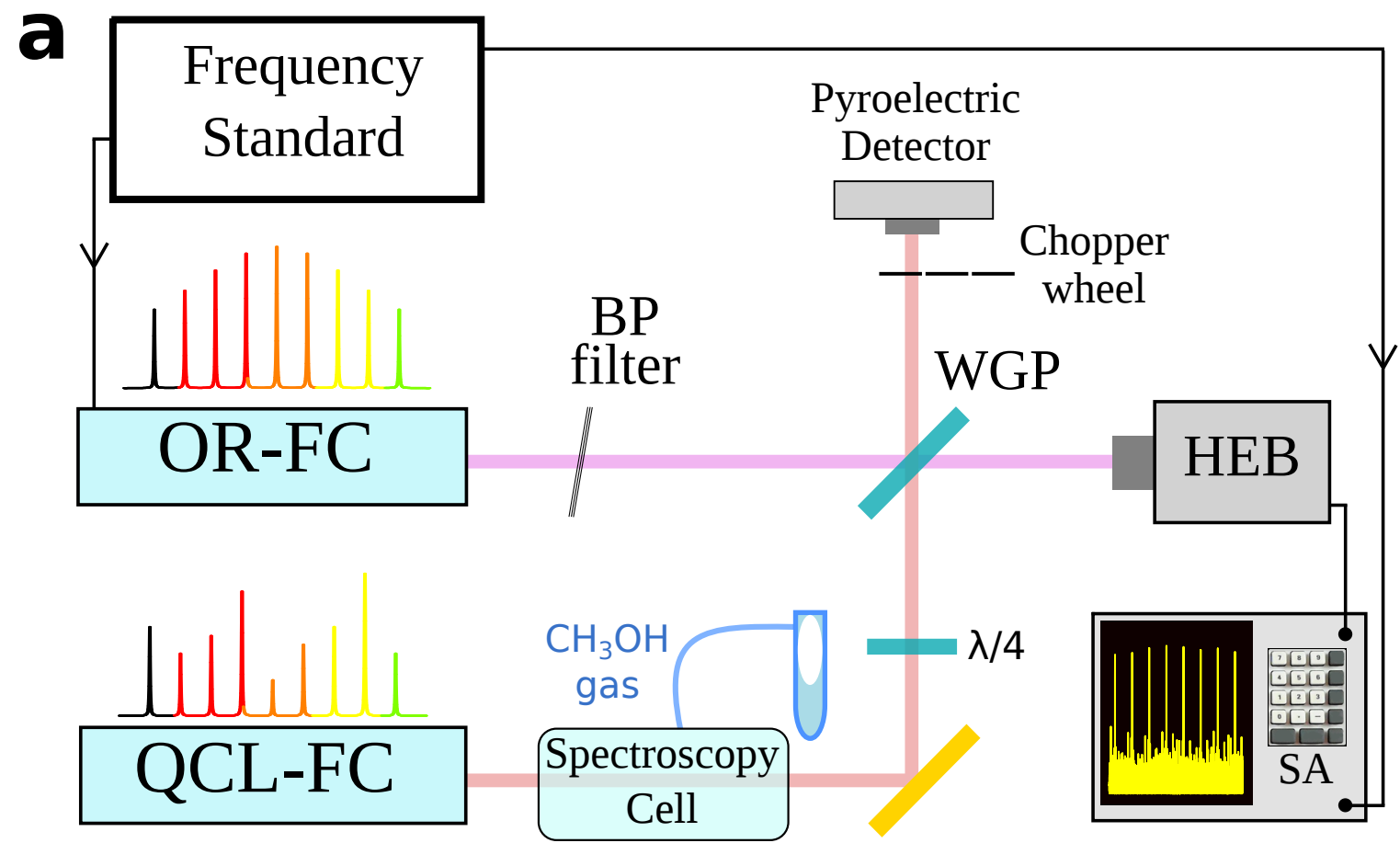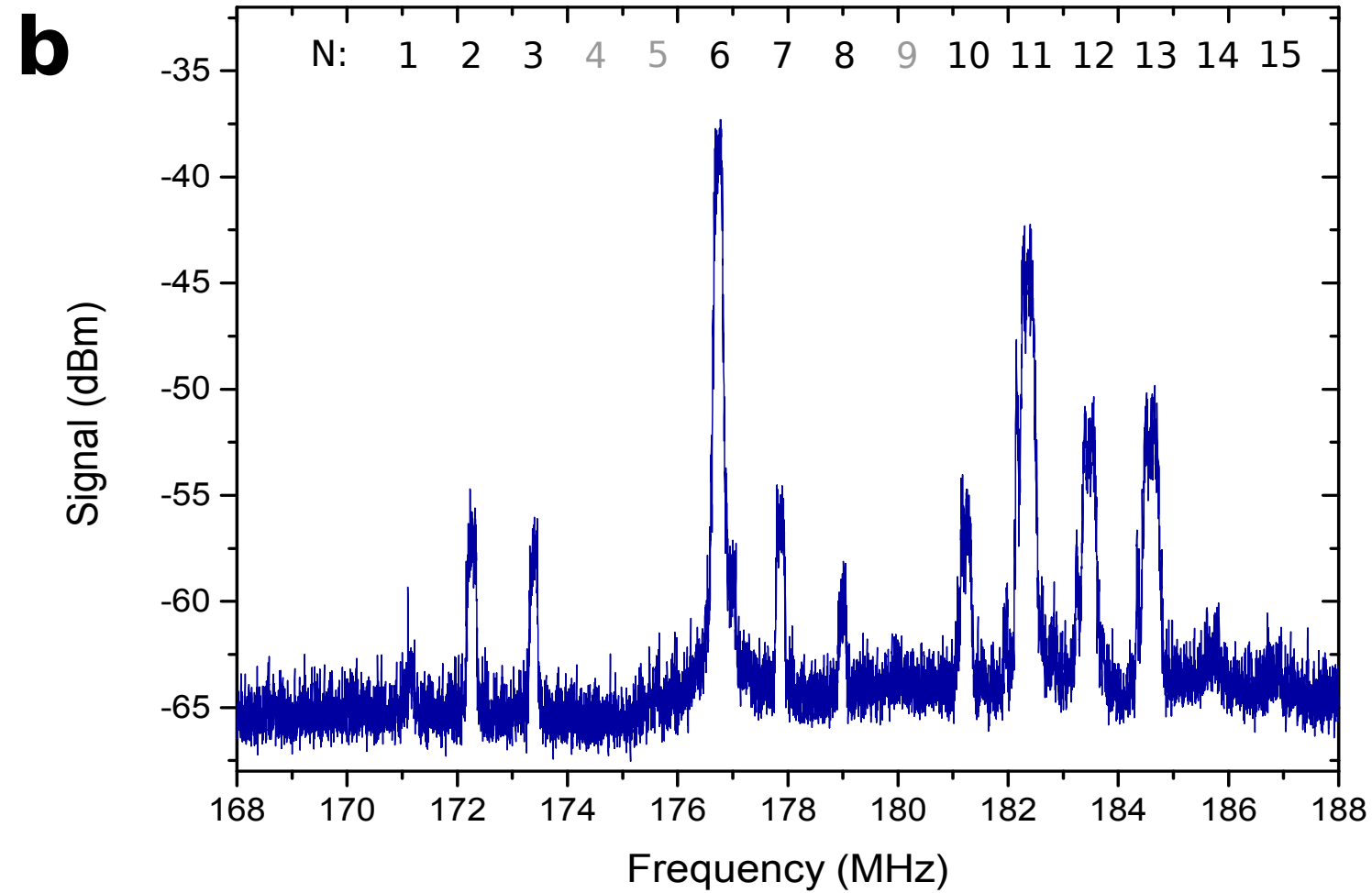

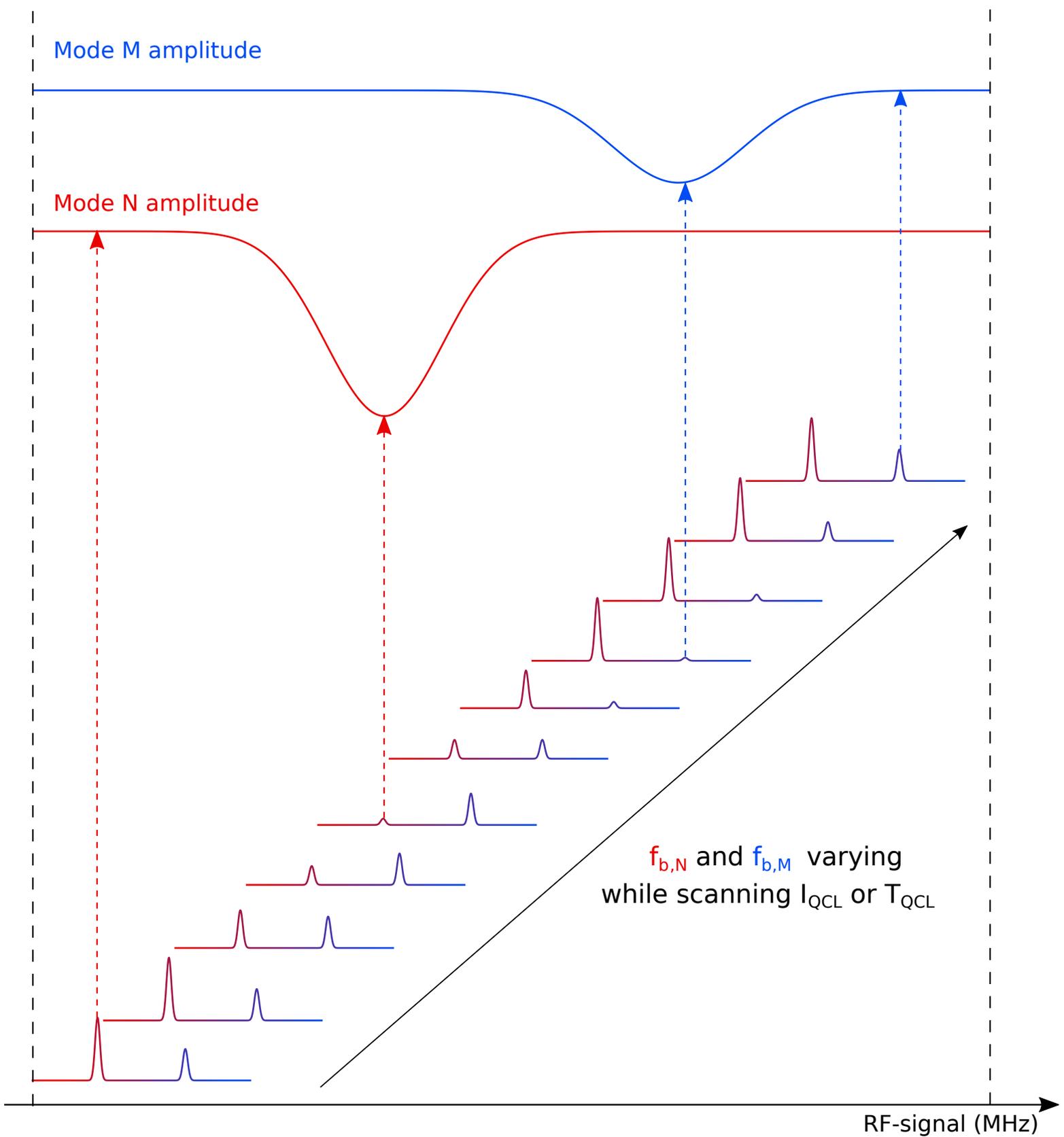

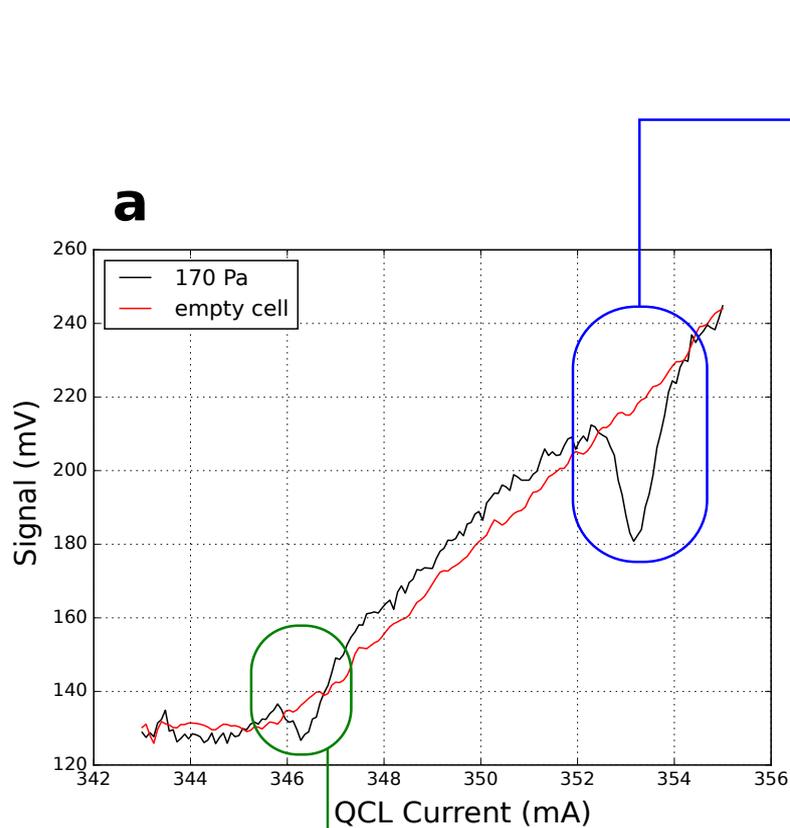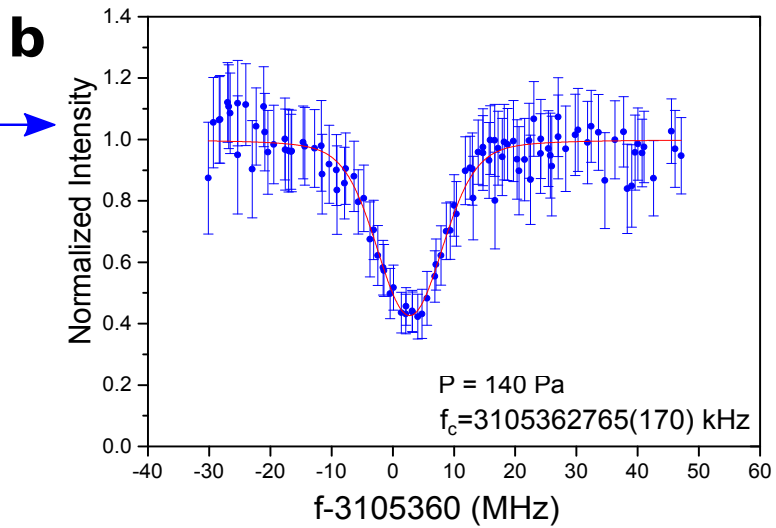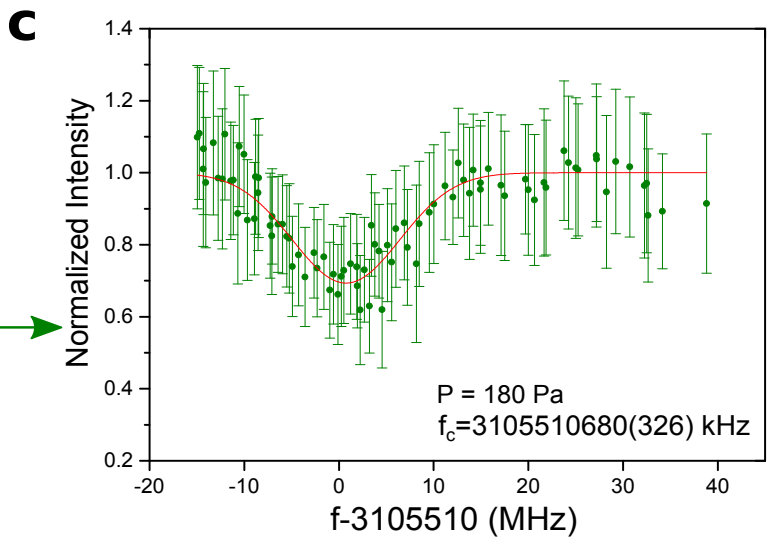

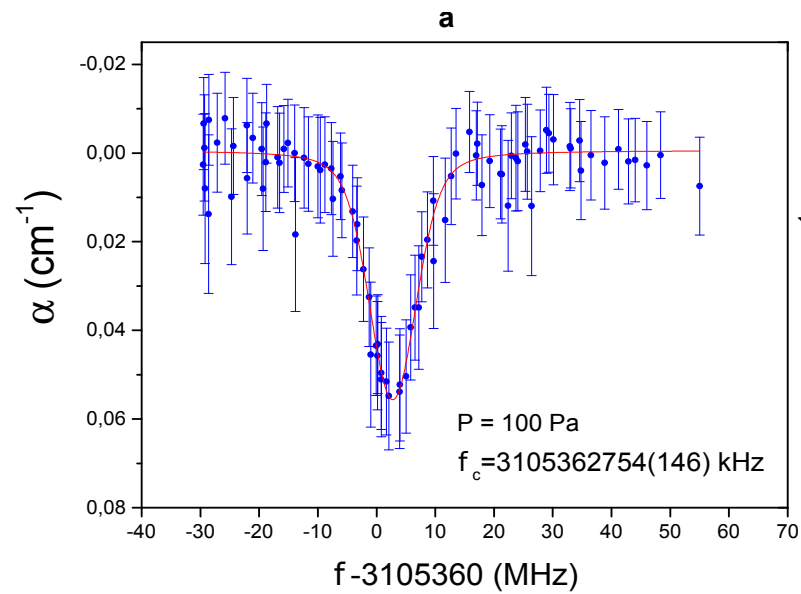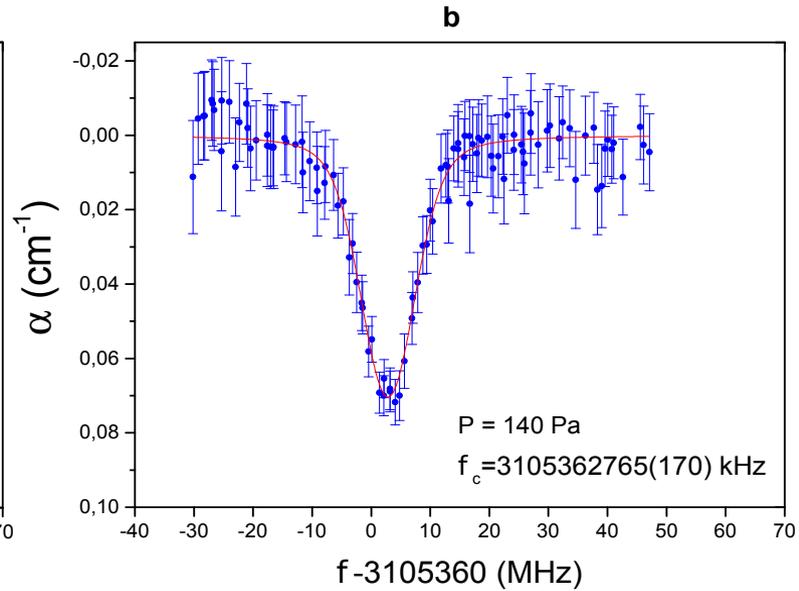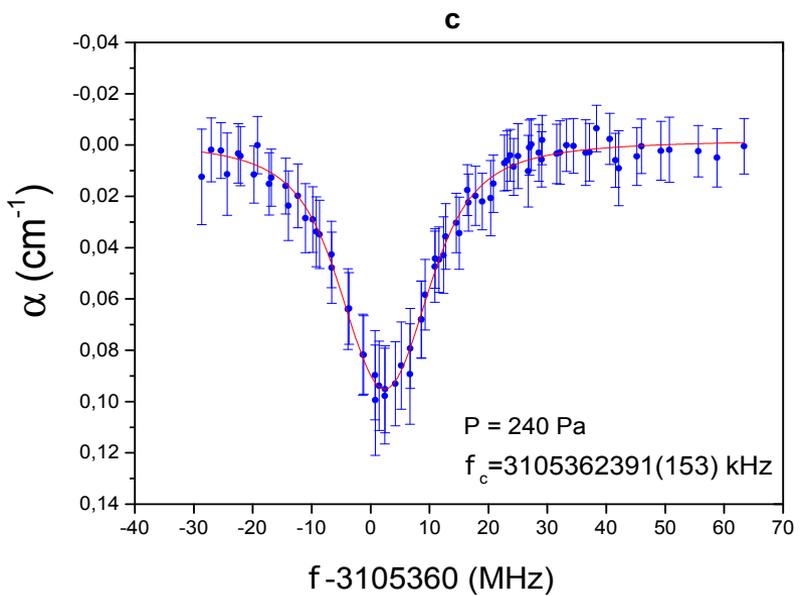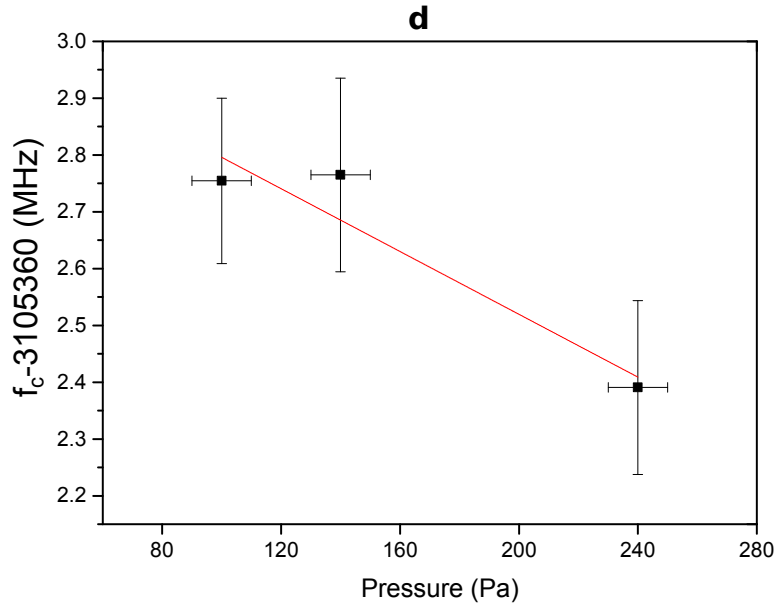

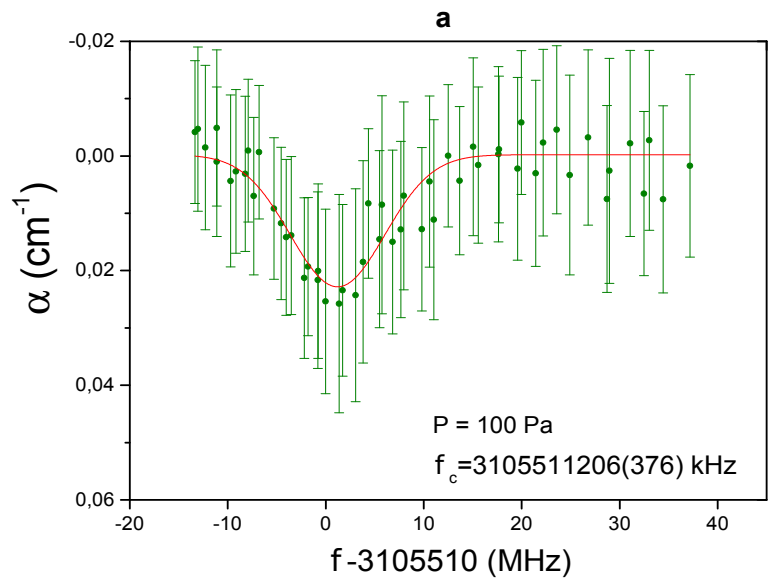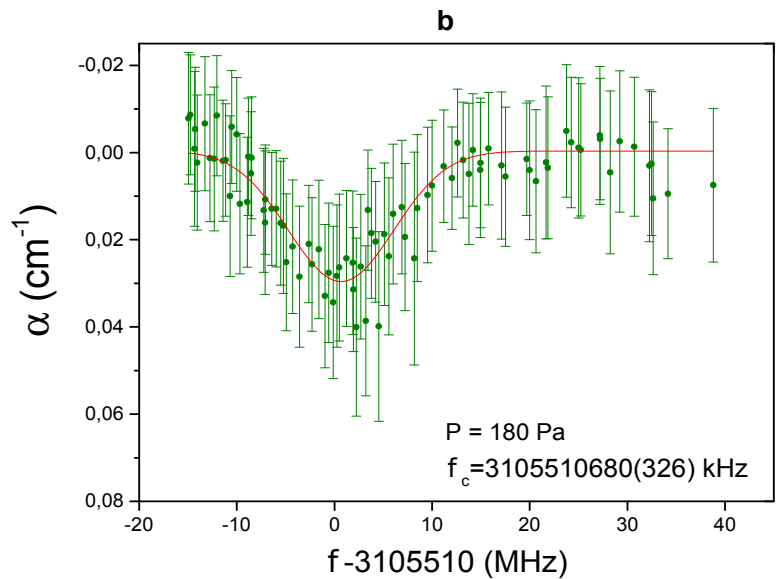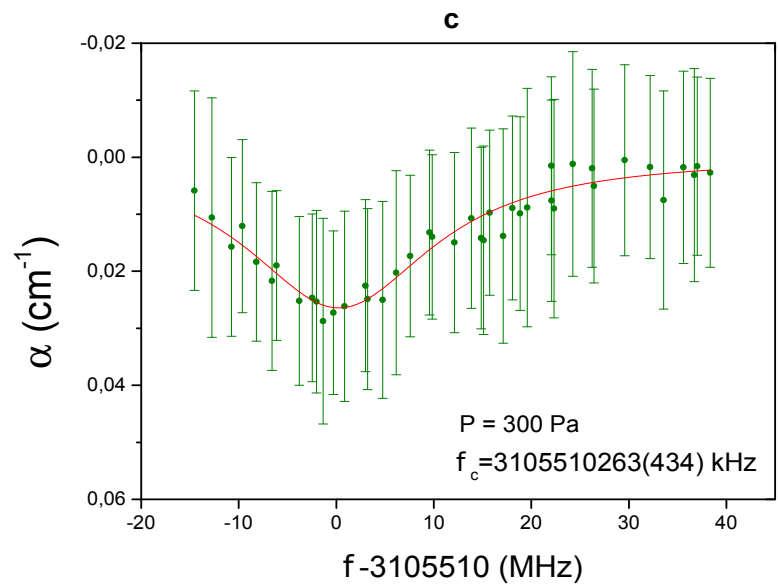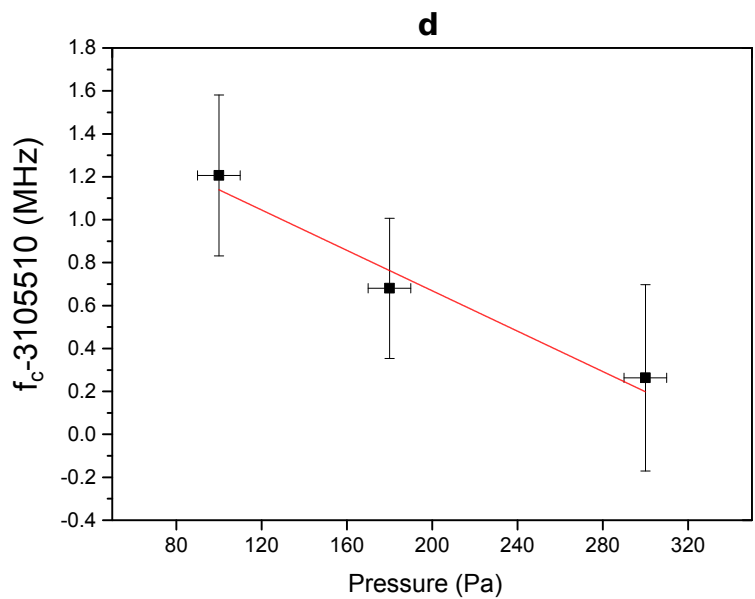

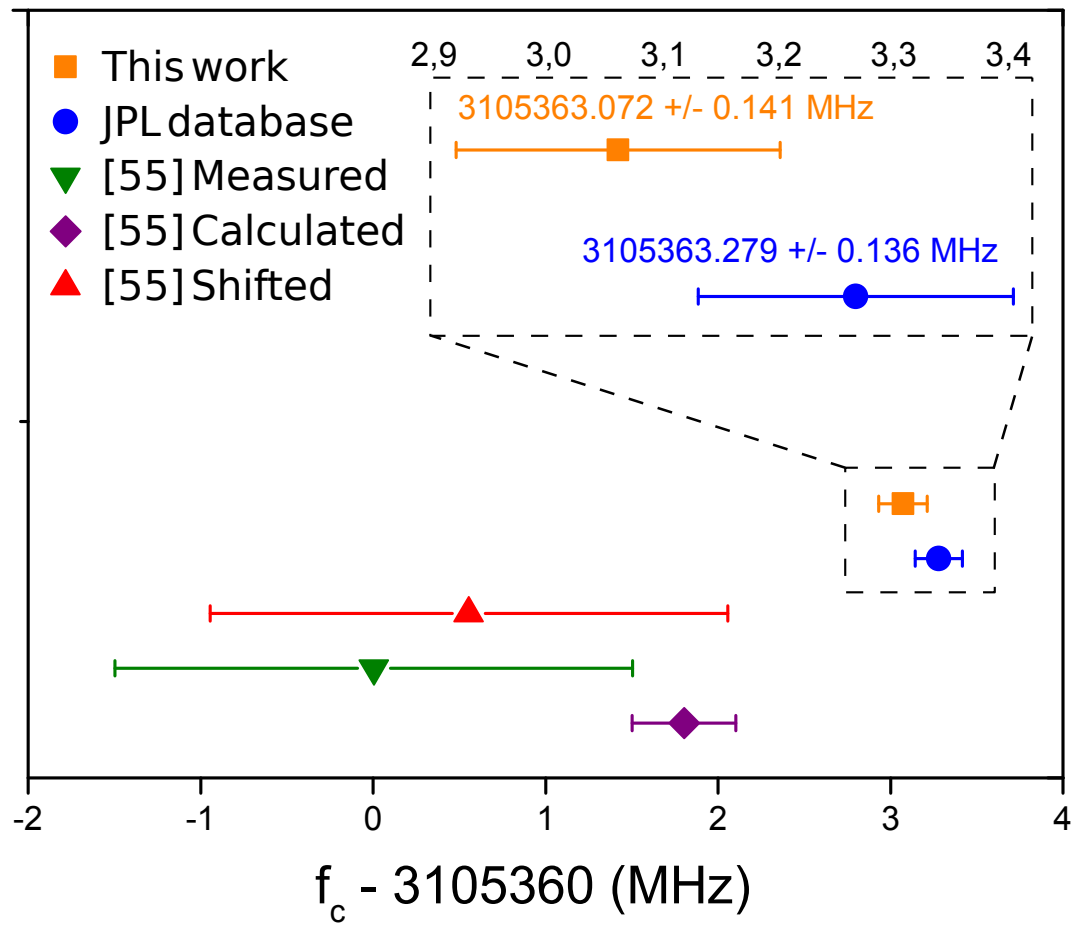
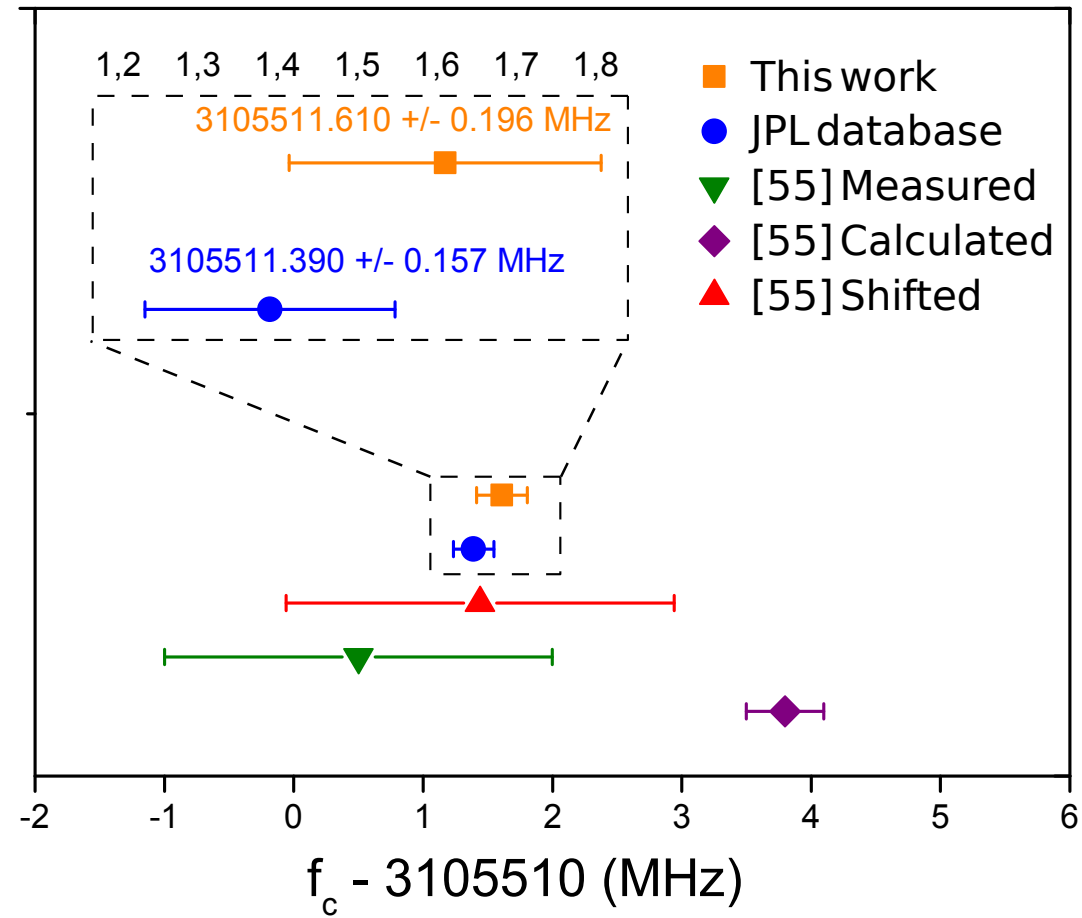

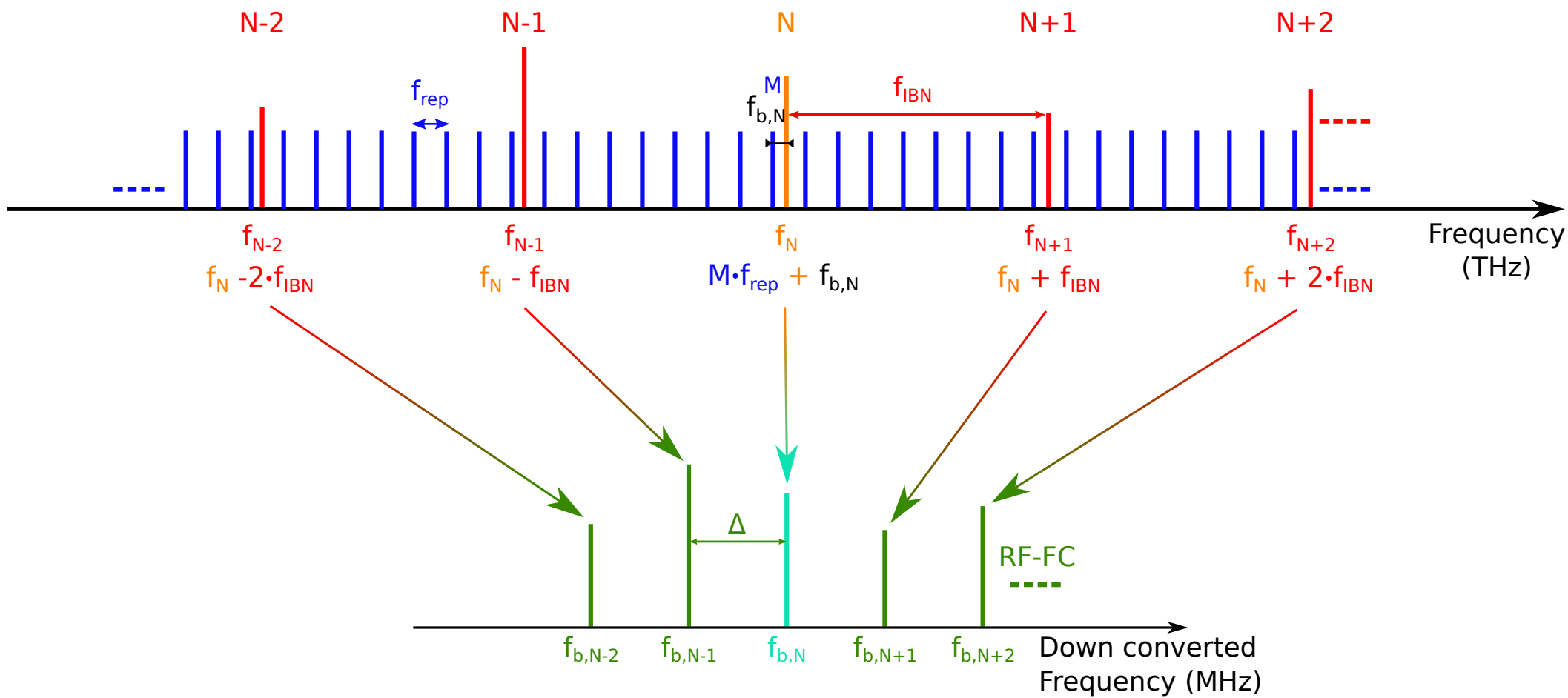

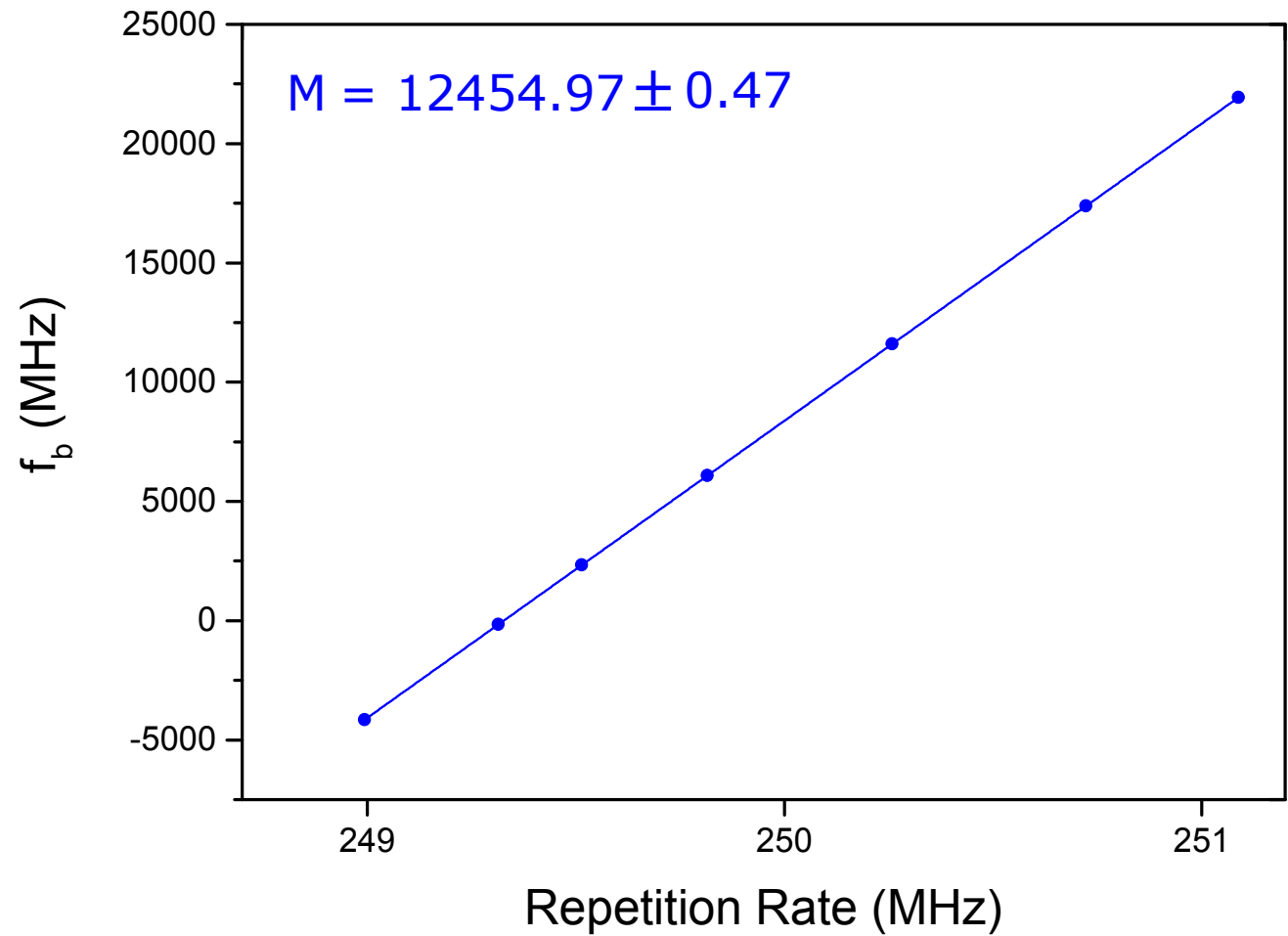 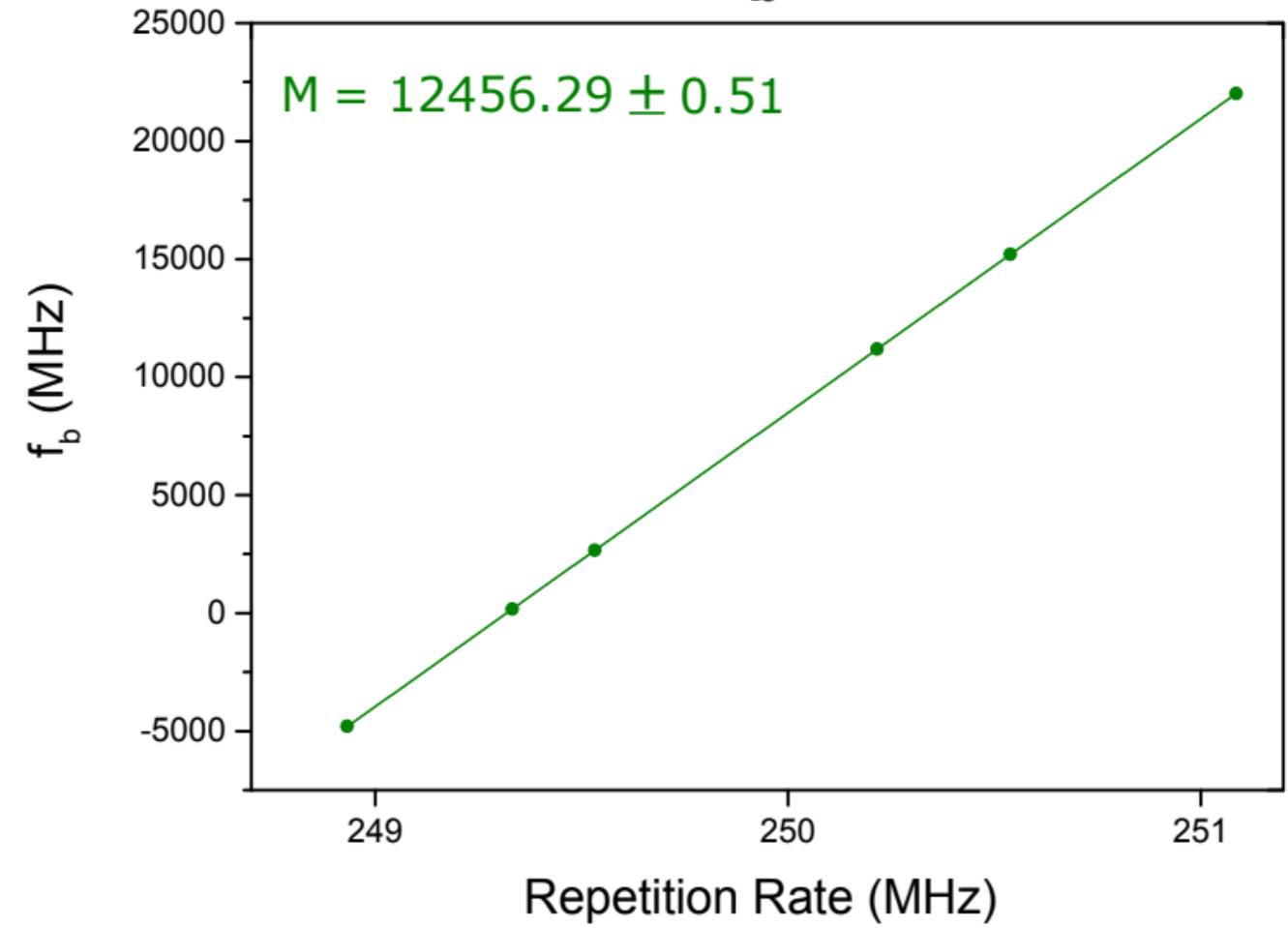

| Transition T1 | | St. error | Transition T2 | | St. error |
|---|---|---|---|---|---|
| Intercept (MHz) | 3.072 | 0.141 | Intercept (MHz) | 1.610 | 0.196 |
| Slope (kHz/Pa) | -2.8 | 0.8 | Slope (kHz/Pa) | -4.7 | 1.0 |